\documentclass[twocolumn,preprintnumbers,superscriptaddress,amsmath,amssymb]{revtex4-1}
\usepackage{graphics,epsfig,subfigure}
\usepackage{ulem}
\usepackage{color}
\usepackage[colorlinks=true, linkcolor=blue, citecolor=blue]{hyperref}

\begin{document}
\newcommand{\PR}[1]{\ensuremath{\left[#1\right]}} 
\newcommand{\PC}[1]{\ensuremath{\left(#1\right)}} 
\newcommand{\PX}[1]{\ensuremath{\left\lbrace#1\right\rbrace}} 
\newcommand{\BR}[1]{\ensuremath{\left\langle#1\right\vert}} 
\newcommand{\KT}[1]{\ensuremath{\left\vert#1\right\rangle}} 
\newcommand{\MD}[1]{\ensuremath{\left\vert#1\right\vert}} 

\title{Cosmological perturbations in the energy-momentum squared gravity theory: constraints from gravitational wave standard sirens and redshift space distortions}
\author{Qi-Ming Fu}\email{fuqiming@snut.edu.cn}
\affiliation{Institute of Physics, Shaanxi University of Technology, Hanzhong 723000, China}
\affiliation{Lanzhou Center for Theoretical Physics, Key Laboratory of Theoretical Physics of Gansu Province, and Key Laboratory of Quantum Theory and Applications of MoE, Lanzhou University, Lanzhou, Gansu 730000, China}

\author{Xin Zhang}\thanks{Corresponding author}\email{zhangxin@mail.neu.edu.cn}
\affiliation{Key Laboratory of Cosmology and Astrophysics (Liaoning) \& College of Sciences, Northeastern University, Shenyang 110819, China}
\affiliation{Key Laboratory of Data Analytics and Optimization for Smart Industry (Ministry of Education), Northeastern University, Shenyang 110819, China}
\affiliation{National Frontiers Science Center for Industrial Intelligence and Systems Optimization, Northeastern University, Shenyang 110819, China}

\begin{abstract}
  We investigate the linear cosmological perturbations in the context of the so-called energy-momentum squared gravity (EMSG) theory. Recent researches show that the EMSG theory can reproduce viable background cosmological evolution comparable to $\Lambda$CDM, while the matter-dominated era exhibits slight distinctions. In this paper, we mainly focus on the power-law EMSG models and derive the equations for the linear cosmological perturbations. We explore the propagation of the gravitational wave (GW) and the growth of matter density perturbation at the first order, and estimate the model parameters from the simulated GW data and the observed redshift space distortion data. Our analysis reveals that the model parameters should be small and positive in $1\sigma$ confidence interval, which indicates that the theory is in good agreement with the observational data and can be regarded as an alternative for the standard cosmological model.
\end{abstract}




\maketitle

\section{Introduction}

Recently, the energy-momentum squared gravity (EMSG) has emerged as a novel approach aimed at addressing the issue of the Big-Bang singularity by introducing the self-coupling of the matter. This theory introduces a scalar, denoted as $\mathbb{T}\equiv T_{\mu\nu}T^{\mu\nu}$, which is constructed from the energy-momentum tensor (EMT) and distinct from theories that rely on the trace of $T_{\mu\nu}$. Their modifications to general relativity (GR) come from the matter instead of the geometry, which results in interesting cosmological outputs. In Ref.~\cite{Roshan044002}, it has been demonstrated that a simple model of the EMSG theory in the metric formulation yields a viable cosmic behavior, presenting a genuine sequence of cosmic epochs. Notably, the key deviation from the standard cosmological picture lies in the absence of an early universe singularity. In essence, the EMSG theory posits the existence of a minimum length and a maximum, yet finite, energy density during the early universe. Presently, interest in this theory is not limited to solve the Big-Bang singularity. Depending on specific EMSG model under consideration, it may propose intriguing modifications to the entire cosmic history, extending beyond the early universe. While the EMSG theory does not evade black hole singularities, it does predict larger masses for neutron stars under ordinary equations of state \cite{Nari024031,Akarsu124017}. For recent work on quark compact stars within the framework of the EMSG theory, readers are referred to Ref.~\cite{Singh100774}.

Various aspects of the EMSG theory have been explored. In Ref.~\cite{Bahamonde083511}, different EMSG models were scrutinized using the dynamical system approach, revealing intriguing cosmological behaviors through the analysis of relevant fixed points. Additional cosmological investigations related to the EMSG theory can be found in Refs.~\cite{Kavuk163,Board123517,Akarsu024011,Akarsu063522,Barbar044058,FariaA127,Akarsu846}. Reference \cite{Kazemi150} investigated in depth the Jeans analysis within the context of the EMSG theory, introducing a novel Jeans mass. The study also explored the local stability of hyper-massive neutron stars by identifying a generalized version of Toomre's parameter in the EMSG theory, outlining constraints on the theory's free parameters. A notable finding reported in Ref.~\cite{Barbar044058} indicates that the bounce mentioned in Ref.~\cite{Roshan044002} does not accurately describe the real universe. Specifically, the EMSG theory exhibits a viable bounce only in a matter-dominated cosmological toy model \cite{Barbar044058}. To achieve a viable bounce, the quadratic scalar $\mathbb{T}$ in the action should be replaced by the term $\mathbb{T}^{5/8}$. In Ref.~\cite{Nazari104026}, the author investigated the motion of light in the weak-field limit of EMSG theory and showed that this theory can pass the solar system tests with flying color. The constraints on the EMSG theory by binary pulsar observations was studied in Ref.~\cite{Nazari044014}. Besides, Ref.~\cite{Akarsu5452} presented an analysis about the gravitational-wave (GW) radiation and radiative behavior of relativistic compact binary systems in the so-called scale-independent EMSG theory. One can see Refs.~\cite{Chen064021,Nazari064016,Rudra100849,Akarsu124059,Chen253,Acquaviva101128,Khodadi101013,Tangphati169149,Akarsu101194} for more investigations on the EMSG theory.

On the other hand, the linear perturbation is an effective method to discriminate different modified gravitational theories since they may have the same cosmological background solution, but the evolutions of the perturbed modes are different. In general, the linear perturbation can be decomposed into scalar, vector, and tensor perturbations. Since the vector mode decays with the cosmological expansion, most works in the literature focus on the scalar and tensor cosmological perturbations. Tensor perturbations correspond to the propagation of GW in the cosmological background, serving as standard sirens for determining the distance-redshift relation \cite{Bertacca32}. Therefore, with the upcoming deployment of ground- and space-based GW detectors like the Einstein Telescope (ET) \cite{Punturo194002}, Laser Interferometer Space Antenna \cite{Seoane5720,Amaro-Seoane00786}, the DECI-hertz Interferometer Gravitational-wave Observatory \cite{Kawamura094011}, the Big-Bang Observer \cite{Harry4887}, Taiji \cite{Wu1844014,Ruan2050075,Hu685,Zhao:2019gyk}, and TianQin \cite{Luo185013,Wang012,Liu103027,Milyukov1067,Mei10332,Fan063016,TianQin:2020hid}, we anticipate the capability to measure cosmological expansion with high precision up to significant redshifts. What's more, the growth rate of matter density perturbation provides an effective approach to estimate the distribution of matter in the universe \cite{Dutta038} and is crucial for theoretically differentiating between various gravitational theories \cite{Li:2015poa,Ishak10122,Khyllep134796,Basilakos212}. The growth of matter density perturbation is highly dependent on the underlying theory of gravity, whether it is GR or a modification of it. Modified field equations result in distinct evolutions of cosmological perturbations, altering the characteristic imprints left in the cosmic microwave background (CMB) and in the matter power spectrum inferred from galaxy clusters. Therefore, it stands out as an exceptional probe for modified gravitational models, facilitating discrimination between these models and conventional dark energy models.

{Numerous studies have extensively explored the viable cosmological solutions within the EMSG theories and the constraints on these models with the cosmological electromagnetic observations, such as, cosmic chronometer \cite{Akarsu024011}, CMB, baryon acoustic oscillations \cite{Akarsu063522}, and type Ia supernovae \cite{FariaA127}, also have been investigated. However, the constraints from the GW signals on the EMSG theories have not been investigated yet and it is interesting to know whether the propagation of GW is affected. Besides, the modifications on the matter stress will affect the growth of matter perturbations. Thus, it is effective to constrain the EMSG theories by using the redshift space distortion (RSD) observations.}

In this paper, we aim to derive the propagation of GW and the linear evolution for the matter density perturbation, and then proceed to extract the observational constraints on the model parameters of some explicit EMSG models, utilizing the mock standard sirens data and the actual RSD data. Besides, although the spherical overdensities within the EMSG theory was investigated in Ref.~\cite{Farsi023524}, the differential equation for the matter density perturbation is derived from the usual continuity equation, which is only valid for some special cases of the EMSG models. In this paper, we start with the non-conserved EMT and obtain the general equations of motion for the matter density perturbation.

This paper is organized as follows. In Sec.~\ref{emsg}, we first give a brief review of the EMSG theory and then present the solutions of two power-law EMSG models. In Sec.~\ref{pert}, we derive the tensor and matter density perturbations of the theory. In Sec.~\ref{const}, we present the formalism to simulate the GW detection from the ET and the method for the RSD data analysis, and then find the fit values of the model parameters by using the mock GW data and the actual RSD data. Section \ref{con} comes with the conclusion.

\section{EMSG cosmological models}~\label{emsg}

Let us start with the action for the EMSG theory \cite{Kavuk163,Roshan044002}:
\begin{eqnarray}
S=\int d^4x \sqrt{-g}(f(R,\mathbb{T})+\mathcal{L}_m),
\end{eqnarray}
where $\mathbb{T}\equiv T_{\mu\nu}T^{\mu\nu}$ and $T_{\mu\nu}$ is the EMT, which represents the same matter with the Lagrangian $\mathcal{L}_m$. Varying the action with respect to the metric $g^{\mu\nu}$, the field equations can be derived as
\begin{eqnarray}
&&f_R R_{\mu\nu}-\frac{1}{2}f g_{\mu\nu}+(g_{\mu\nu}\nabla_{\tau}\nabla^{\tau}-\nabla_{\mu}\nabla_{\nu})f_R \nonumber\\
&&=\frac{1}{2}T_{\mu\nu}-f_{\mathbb{T}}\theta_{\mu\nu}, ~\label{eeom}
\end{eqnarray}
where $f_R\equiv\frac{\partial f}{\partial R}$, $f_{\mathbb{T}}\equiv\frac{\partial f}{\partial \mathbb{T}}$, and
\begin{eqnarray}
\theta_{\mu\nu}&=&-2\mathcal{L}_m\left(T_{\mu\nu}-\frac{1}{2}g_{\mu\nu}T\right)-T T_{\mu\nu}+2T_{\mu}^{\alpha}T_{\nu\alpha} \nonumber\\
               &-&4T^{\alpha\beta}\frac{\partial^2 \mathcal{L}_m}{\partial g^{\mu\nu}\partial g^{\alpha\beta}}.
\end{eqnarray}
The matter Lagrangian giving the perfect-fluid EMT is not unique, since both of $\mathcal{L}_m=p$ and $\mathcal{L}_m=-\rho$ can provide the same EMT. In this paper, we consider $\mathcal{L}_m = p$ for consistency. Then, the tensor $\theta_{\mu\nu}$ reduces to
\begin{eqnarray}
\theta_{\mu\nu}=-2\mathcal{L}_m\left(T_{\mu\nu}-\frac{1}{2}g_{\mu\nu}T\right)-T T_{\mu\nu}+2T_{\mu}^{\alpha}T_{\nu\alpha}.
\end{eqnarray}

The covariant divergence of Eq.~(\ref{eeom}) is
\begin{eqnarray}
\nabla^{\mu}T_{\mu\nu}=-f_{\mathbb{T}}g_{\mu\nu}\nabla^{\mu}\mathbb{T}+2\nabla^{\mu}(f_{\mathbb{T}}\theta_{\mu\nu}). ~\label{cEMT}
\end{eqnarray}
The nonvanishing of the right-hand side of the above equation indicates explicitly that the covariant conservation of the EMT is not a priori in the EMSG theory. These nonvanishing terms act as source terms accounting for the stress-energy transfer between matter and geometry. In order to proceed on, we need to specify explicit models of this theory. Without loss of generality, we will focus on the power-law one without the mixed terms between $R$ and $\mathbb{T}$, which is one of the most viable models for cosmology \cite{Kavuk163,Roshan044002,Board123517,Akarsu024011,Akarsu063522,FariaA127,Barbar044058,Farsi023524}, although our methodology can be applied for any functional form of $f(R, \mathbb{T})$. Besides, the matter density perturbation will be considerably simplified without the mixed terms.
The power-law scenario corresponds to \cite{Board123517,Akarsu024011}:
\begin{eqnarray}
f(R, \mathbb{T})=\frac{1}{2\kappa}R+\eta \mathbb{T}^n,
\end{eqnarray}
where $\kappa=8\pi G$ with $G$ the Newton's constant, $\eta$ is the coupling constant signifying the strength of the EMT-powered modification to gravity, and $n$ is the power of $\mathbb{T}$. In general, this system can only be solved numerically. However, analytical solutions can be found for some particular values of $n$, i.e., $n=0$ \cite{Akarsu024011}, $n=1/2$ \cite{Kavuk163}, and $n=1$ \cite{Roshan044002}. Obviously, for $n=0$, this theory reduces to GR, which is the trivial case. The scenario with $n=1/2$ corresponds to the so-called scale-independent EMSG theory \cite{Akarsu063522}, which exhibits the potential to be influential across various cosmological epochs characterized by different energy density scales, and plays a significant role in both early and late-time evolutions of the universe. The model with $n=1$ corresponds to the original EMSG theory, which can pass the solar-system weak-field tests \cite{Nazari104026,Akarsu101305}. In the following, we will take $n=1/2$ and $n=1$ as two examples to investigate the linear cosmological perturbations of the EMSG theory.

For $n=1$, the covariant conservation of the EMT can be accomplished and the usual continuity equation is satisfied, which leads to the solution for the matter density $\rho=\rho_0 a^{-3}$. The Friedmann equation in terms of the redshift can be expressed as \cite{Roshan044002}
\begin{eqnarray}
H^2=H_0^2\big(\Omega_m(1+z)^3+\Omega_Q(1+z)^6+\Omega_{\Lambda}\big),
\end{eqnarray}
where $\Omega_m\equiv\frac{\kappa \rho_0}{3H_0^2}$, $\Omega_{\Lambda}\equiv\frac{\Lambda}{3H_0^2}$, and $\Omega_Q\equiv\frac{\eta \rho_0^2}{6H_0^2}=\frac{3\eta H_0^2\Omega_m^2}{2\kappa^2}$, which satisfy $\Omega_m+\Omega_Q+\Omega_{\Lambda}=1$.

For $n=1/2$, Eq.~(\ref{cEMT}) leads to $\rho=\rho_0 a^{-3(1+\eta_t)}$ with $\eta_t\equiv\frac{\eta}{2\kappa}$, and the Friedmann equation is given by \cite{Kavuk163}
\begin{eqnarray}
H^2=H_0^2\left(\Omega_m(1+z)^{3\left(1+\eta_t\right)}+1-\Omega_m\right).
\end{eqnarray}

\section{perturbations}~\label{pert}

In this section we will establish the key equations governing the linear cosmological perturbations of the gravitational field in the EMSG theory. These linear perturbations will be assumed to evolve on a homogeneous and isotropic background described by the Friedmann-Lema\^{\i}tre-Robertson-Walker metric. According to the $SO(3)$ symmetry of the background, these small perturbations can be decomposed into scalar, vector, and tensor perturbations, which can be studied independently. Since the vector mode decays with the cosmological expansion,
we mainly restrict our attention to the tensor and scalar modes of the cosmological perturbations.

\subsection{Tensor perturbation}

Let us focus on the equations of motion of GW first. The perturbed metric can be written as
\begin{eqnarray}
ds^2=-dt^2+a(t)^2(\eta_{ij}+h_{ij})dx^i dx^j,
\end{eqnarray}
where $h_{ij}$ represents the GW perturbation, which satisfies the transverse-traceless conditions, i.e., $\eta^{ij}h_{ij}=0$ and $\partial^i h_{ij}=0$. Then, the field equations for $h_{ij}$ derived from Eq.~(\ref{eeom}) read
\begin{eqnarray}
\ddot{h}_{ij}+3H\left(1-\beta\right)\dot{h}_{ij}-a^{-2}\eta^{kl}\partial_k\partial_l h_{ij}=0, ~\label{tp1}
\end{eqnarray}
where the dimensionless parameter $\beta$ is defined as $\beta\equiv -\frac{\dot{f}_R}{3H f_R}$ and the dot standards for the derivative with respect to $t$. Equation (\ref{tp1}) can be rewritten in terms of the conformal time as follows
\begin{eqnarray}
\partial_{\tau}^2 h_{ij}+2H\left(1-\frac{3}{2}\beta\right)\partial_{\tau}h_{ij}-\eta^{kl}\partial_k\partial_l h_{ij}=0,
\end{eqnarray}
where $\tau$ is the conformal time defined by $d\tau=dt/a(t)$. The above equation indicates that the speed of GW is equal to the speed of light, and thus the constraints from the observation of GW170817/GRB170817A is trivially satisfied in the EMSG theory \cite{AbbottL13}. Since the presence of the friction term, the GW luminosity distance is given by \cite{Belgacem023510,Belgacem024}
\begin{eqnarray}
d_L^{GW}=d_L^{EM}\text{exp}\left[-\frac{3}{2}\int_0^z \frac{dz'}{1+z'}\beta(z')\right].
\end{eqnarray}
Here, $d_L^{EM}$ is the standard luminosity distance for an electromagnetic signal, namely
\begin{eqnarray}
d_L^{EM}=(1+z)\int_0^z \frac{dz'}{H(z')},
\end{eqnarray}
which is the same for the gravitational radiation in GR.

\subsection{Matter density perturbation}

Next, let us consider the evolution of the matter density perturbation for the EMSG theory. In the Newtonian gauge, the metric of the linear scalar perturbations can be written as
\begin{eqnarray}
ds^2=-(1+2\Phi)dt^2+a(t)^2(1-2\Psi)\eta_{ij}dx^i dx^j, ~\label{spm}
\end{eqnarray}
where $\Phi(t, x_i)$ and $\Psi(t, x_i)$ represent the scalar perturbations. The components of the perturbed EMT can be given by
\begin{eqnarray}
\delta T_0^0&=&-\delta\rho \equiv -\rho\delta, \\
\delta T_i^0&=&\rho \partial_i v, \quad \delta T_j^i=0,
\end{eqnarray}
where we have assumed the universe at the time where we are performing the perturbations is in the dust dominated phase. By considering Eqs.~(\ref{eeom}) and (\ref{spm}), one can obtain the perturbed equations with the components $(00),~(ij),~(ii)$, and $(0i)=(i0)$, where $i,~j=1,2,3$, $i\neq j$ in Fourier space, respectively given by
\begin{eqnarray}
&&-6 H^2 \Phi -6 H \dot{\Psi}-\frac{2 k^2 \Psi }{a^2}-\frac{\delta \rho }{2 f_R}+\frac{1}{f_R}\bigg(3 H \delta \dot{f}_R \nonumber\\
&&-6 H \Phi  \dot{f}_R-3 \dot{\Psi} \dot{f}_R-  \rho \delta \rho  \left(f_{\mathbb{T}}+2 \rho ^2 f_{\mathbb{T}\mathbb{T}}\right)+\delta f_R\Big(\frac{k^2}{a^2} \nonumber\\
&&-3 H^2-3 \dot{H}\Big)\bigg)=0, ~\label{00} \\
&&\Phi-\Psi=-\frac{f_{RR}}{f_R}\delta R, ~\label{12} \\
&&2 f_R \left(2 \dot{H} (\Phi +\Psi )+H \dot{\Phi}+3 H \left(H \Phi +\dot{\Psi}\right)+\ddot{\Psi}\right) \nonumber\\
&&+4 H \Phi  \dot{f}_R-2 H \Psi  \dot{f}_R+\dot{\Phi} \dot{f}_R+2 (\Phi +\Psi ) \ddot{f}_R \nonumber\\
&&+2 \dot{\Psi} \dot{f}_R+\rho  f_{\mathbb{T}} (2 \rho  \Psi -\delta \rho )+\left(3 H^2+\dot{H}\right) \delta f_R \nonumber\\
&&-2 H \delta \dot{f}_R+\rho  \Psi -\delta \ddot{f}_R=0, \\
&&\Phi  \left(2 H f_R+\dot{f}_R\right)+H \delta f_R+2 f_R \dot{\Psi}-\dot{f}_R \nonumber\\
&&+\frac{1}{2} \rho  v \left(1+2 \rho  f_{\mathbb{T}}\right)=0,
\end{eqnarray}
with
\begin{eqnarray}
\delta R &=& -\frac{2}{a^2} \Big(3 a^2 \left(2 \Phi  \dot{H}+H \dot{\Phi}+4 H \left(H \Phi +\dot{\Psi}\right)+\ddot{\Psi}\right) \nonumber\\
&-&k^2 (\Phi-2 \Psi)\Big). ~\label{deltaR}
\end{eqnarray}
It should be stressed again that the above perturbed equations ignored the mixed terms between $R$ and $\mathbb{T}$ since we only focus on the power-law model. Besides, one can easily notice that by taking $f(R, \mathbb{T})=f(R)$ in the above results, the perturbed equations in $f(R)$ gravity are recovered \cite{Bean064020}. The covariant divergence of the EMT renders to the following first-order equations
\begin{eqnarray}
&&\left(1+2 \rho  f_{\mathbb{T}}+4 \rho ^3 f_{\mathbb{T}\mathbb{T}}\right)\delta \dot{\rho}+ \bigg(3 H+2 f_{\mathbb{T}} \left(6 H \rho +\dot{\rho}\right) \nonumber\\
&&+4 \rho ^2 f_{\mathbb{T}\mathbb{T}} \left(3 H \rho +2 \dot{\rho}\right)+4 \rho  \dot{f}_{\mathbb{T}}+4 \rho ^3 \dot{f}_{\mathbb{T}\mathbb{T}}\bigg)\delta \rho= \nonumber\\
&&\frac{k^2 \rho  v \left(1+2 \rho  f_{\mathbb{T}}\right)}{a^2}+3 \rho  \dot{\Psi} \left(1+2 \rho  f_{\mathbb{T}}\right), ~\label{c1} \\
&&\left(2 \rho  f_{\mathbb{T}}+1\right) \left(\dot{v}+\Phi \right)+\bigg(2 \rho  \dot{f}_{\mathbb{T}}
+2  f_{\mathbb{T}} \left(3 H \rho +2 \dot{\rho}\right) \nonumber\\
&&+3 H+\frac{\dot{\rho}}{\rho}\bigg)v+2 \delta  \rho  f_{\mathbb{T}}=0. ~\label{c2}
\end{eqnarray}
for the temporal and spatial components, respectively.

From Eqs.~(\ref{00}), (\ref{12}) and (\ref{deltaR}), and considering the subhorizon approximation, which is given by $H^2\Psi,~H\dot{\Psi},~\ddot{\Psi}\ll k^2\Psi/a^2$, one can obtain
\begin{eqnarray}
\Psi \!\!&=&\!\!\frac{1}{2} \Phi  \left(1+\frac{a^2 f_R}{a^2 f_R+4 k^2 f_{RR}}\right), \\
\Phi \!\!&=&\!\!-\frac{\delta \rho  \left(1+2 \rho  f_{\mathbb{T}}+4 \rho ^3 f_{\mathbb{T}\mathbb{T}}\right) \left(a^4 f_R+4 a^2 k^2 f_{RR}\right)}{4 k^2 f_R \left(a^2 f_R+3 k^2 f_{RR}\right)},~~~
\end{eqnarray}
Then combing Eqs.~(\ref{c1}) and (\ref{c2}) and inserting the last two expressions, the equation of motion for the matter density perturbation ($\delta_m\equiv \delta\rho/\rho$) can be derived as
\begin{eqnarray}
\ddot{\delta}_m+A(t)\dot{\delta}_m+B(t)\delta_m=0, ~\label{deltaeq}
\end{eqnarray}
where $A(t)$ and $B(t)$ are presented in the Appendix, and this differential equation can be solved by setting the initial conditions in the deep matter dominated era \cite{Tsujikawa023514,Nesseris023542}. In the following, we will transform this equation to redshift coordinates and analyze the dynamics of the matter density perturbation as a function of the redshift.

In order to compare the model with the observational data, we use the data on $f\sigma_8$ \cite{Kazantzidis103503,Shahidi274,Shahidi084033} defined as
\begin{eqnarray}
f\sigma_8(z)=\sigma_8(z)\frac{d \text{ln}\delta_m(z)}{d \text{ln}a}, \quad \sigma_8(z)=\sigma_8\frac{\delta_m(z)}{\delta_m(0)},
\end{eqnarray}
where $\sigma_8\equiv\sigma_8(0)$ stands for the current value. The comoving wavenumber is set to $k=0.1h$ Mpc$^{-1}$ to ensure that we are dealing with the linear regime \cite{Zhang2244,Basilakos123529}, and $h$ represents the reduced Hubble constant satisfying $H_0=100 h$ km/s/Mpc. In the following, we will estimate the fit values of the parameters $H_0$, $\Omega_{m}\equiv \Omega_{m}(0)$, $\sigma_8$, and $\eta_t$ or $\Omega_Q$ with the aid of the simulated GW data and the observed RSD data.

\section{Methodology and constraints from Standard sirens and redshift space distortion}~\label{const}

In this section, we first briefly review the simulation of GW detections from the ET. Then, we present the settings for the RSD data analysis. Finally, we explore the constraints on the model parameters through a Markov Chain Monte Carlo (MCMC) analysis by using the data from the simulated GW detections and the observed RSD data sets.

\subsection{Simulation for the gravitational wave detections}

Following Refs.~\cite{Sathyaprakash215006,Zhao023005}, the masses of neutron star (NS) and black hole (BH) are assumed to be uniformly distributed in the range $[1, 2] M_{\odot}$ and $[3, 10] M_{\odot}$, respectively. Here, $M_{\odot}$ is the solar mass. The ratio between NS-NS and BH-NS events is taken to be $0.03$ according to the prediction of the Advanced LIGO-Virgo network \cite{Abadie223}. The redshift distribution of the sources takes the form
\begin{eqnarray}
P(z)\propto \frac{4\pi d_C^2(z)R(z)}{H(z)(1+z)},
\end{eqnarray}
where $d_C$ stands for the comoving distance, which is defined as $d_C(z)\equiv \int_0^z 1/H(z')dz'$, and $R(z)$ denotes the time evolution of the burst rate and takes the form
\begin{eqnarray}
R(z)=\left\{
       \begin{array}{ll}
         1+2z, & \hbox{$z \leqslant 1$} \\
         \frac{3}{4}(5-z), & \hbox{$1 < z <5$} \\
         0, & \hbox{$z \geqslant 5$.}
       \end{array}
     \right.
\end{eqnarray}
The strain of GWs in GW interferometers can be expressed as \cite{Cai044024}
\begin{eqnarray}
h(t)=F_+(\theta,\phi,\psi)h_+(t)+F_{\times}(\theta,\phi,\psi)h_{\times}(t),
\end{eqnarray}
where $\psi$ is the polarization angle and ($\theta,\phi$) are the angles describing the location of the source in the sky, and $F_+$ and $F_{\times}$ represent the detector antenna pattern functions of the ET \cite{Zhao023005}, which are respectively given by
\begin{eqnarray}
F_+^{(1)}(\theta,\phi,\psi)&=&\frac{\sqrt{3}}{2}\bigg[\frac{1}{2}\big(1+\text{cos}^2(\theta)\big)\text{cos}(2\phi)\text{cos}(2\psi) \nonumber\\
                           &-&\text{cos}(\theta)\text{sin}(2\phi)\text{sin}(2\psi)\bigg], \\
F_{\times}^{(1)}(\theta,\phi,\psi)&=&\frac{\sqrt{3}}{2}\bigg[\frac{1}{2}\big(1+\text{cos}^2(\theta)\big)\text{cos}(2\phi)\text{sin}(2\psi) \nonumber\\
                           &+&\text{cos}(\theta)\text{sin}(2\phi)\text{cos}(2\psi)\bigg].
\end{eqnarray}
Since the three interferometers are arranged in an equilateral triangle, the other two antenna pattern functions can be expressed as $F_{+,\times}^{(2)}(\theta,\phi,\psi)=F_{+,\times}^{(1)}(\theta,\phi+2\pi/3,\psi)$ and $F_{+,\times}^{(3)}(\theta,\phi,\psi)=F_{+,\times}^{(1)}(\theta,\phi+4\pi/3,\psi)$, respectively.

By applying the stationary phase approximation, the Fourier transform $\mathcal{H}(f)$ of the strain $h(t)$ is \cite{Zhao023005,Li}
\begin{eqnarray}
\mathcal{H}(f)=\mathcal{A}f^{-7/6}\text{exp}(i\Psi(f)),
\end{eqnarray}
with $\Psi(f)$ a phase and the Fourier amplitude given by
\begin{eqnarray}
\mathcal{A}&=&\frac{1}{d_L}\sqrt{F_+^2\big(1+\text{cos}^2(\iota)\big)^2+4F_{\times}^2\text{cos}^2(\iota)} \nonumber\\
           &\times& \sqrt{\frac{5\pi}{96}}\pi^{-7/6}\mathcal{M}_c^{5/6}, ~\label{amp}
\end{eqnarray}
where $\mathcal{M}_c$ is the chirp mass defined by $\mathcal{M}_c=M\eta^{3/5}$, $M=m_1+m_2$ is the total mass of the coalescence of the binary system with component masses $m_1$ and $m_2$, and $\eta=m_1 m_2/M^2$ is the symmetric mass ratio. In the observer frame, the observed chirp mass is related to the physical chirp mass by a redshift factor, i.e., $\mathcal{M}_{c,\text{obs}}  = (1 + z)\mathcal{M}_{c,\text{phys}}$.
Besides, $\mathcal{M}_c$ in Eq.~(\ref{amp}) denotes the observed chrip mass, while $\iota$ represents the angle of inclination of the binary's orbital angular momentum relative to the line of sight. The definitions for the phase $\Psi(f)$ can be found in Refs.~\cite{Zhao023005,Li,Blanchet024004,Sathyaprakash2}. Given the expected strong beaming of short gamma-ray bursts (SGRBs), the coincident observations of SGRBs suggest that the binaries are predominantly face-on oriented, meaning $\iota \simeq 0$, with the maximum inclination being approximately $\iota = 20^{\circ}$. Notably, averaging the Fisher matrix over the inclination $\iota$ and the polarization $\psi$, subject to the constraint $\iota < 20^{\circ}$, is nearly equivalent to assuming $\iota=0$ in the simulation \cite{Li}. Consequently, we can simplify the simulation of GW sources by setting $\iota=0$.

The performance of a GW detector is characterized by the one-sided noise power spectral density $S_h(f)$ (PSD). We adopt the noise PSD of the ET from Ref.~\cite{Zhao023005}. The combined signal-to-noise ratio for a network of three independent interferometers is given by
\begin{eqnarray}
\rho=\sqrt{\sum^3_{i=1}\big(\rho^{(i)}\big)^2},
\end{eqnarray}
where $\rho^{(i)}=\sqrt{\langle\mathcal{H}^{(i)},\mathcal{H}^{(i)}\rangle}$. The inner product of two functions $a(t)$ and $b(t)$ is defined as
\begin{eqnarray}
\langle a, b\rangle = 4\int_{f_{\text{lower}}}^{f_{\text{upper}}}\frac{\tilde{a}(f)\tilde{b}^*(f)+\tilde{a}^*\tilde{b}(f)}{2}\frac{df}{S_h(f)},
\end{eqnarray}
where $\tilde{a}(f)$ and $\tilde{b}(f)$ are the Fourier transforms of $a(t)$ and $b(t)$, respectively. The lower limit in frequency for ET is $f_{\text{lower}}=1$ Hz, and the upper limit is given by $f_{\text{upper}} = 2/(6^{3/2}2\pi M_{\text{obs}})$, where $M_{\text{obs}} = (1 + z)M_{\text{phys}}$ is the observed total mass.

The standard Fisher matrix method is employed to estimate the instrumental error in luminosity distance, assuming that this parameter is uncorrelated with any other GW parameters \cite{Zhao023005,Li}. This results in
\begin{eqnarray}
\sigma_{d_L}^{\text{inst}}\simeq \sqrt{\left\langle\frac{\partial\mathcal{H}}{\partial d_L}, \frac{\partial\mathcal{H}}{\partial d_L}\right\rangle^{-1}}.
\end{eqnarray}
Since $\mathcal{H}\propto d_L^{-1}$, we have $\sigma_{d_L}^{\text{inst}}\simeq d_L/\rho$. To account for the effect of inclination $\iota$, where $0^{\circ}<\iota<90^{\circ}$, a factor of $2$ is added to the instrumental error, resulting in
\begin{eqnarray}
\sigma_{d_L}^{\text{inst}} \simeq \frac{2d_L}{\rho}.
\end{eqnarray}
An additional error due to gravitational lensing must be considered. For ET, this error is $\sigma_{d_L}^{\text{inst}}=0.05 z d_L$. Thus, the total uncertainty on luminosity distance is given by
\begin{eqnarray}
\sigma_{d_L}&=&\sqrt{\big(\sigma_{d_L}^{\text{inst}}\big)^2+\big(\sigma_{d_L}^{\text{inst}}\big)^2} \nonumber\\
            &=&\sqrt{\left(\frac{2d_L}{\rho}\right)^2+\big(0.05 z d_L\big)^2}.
\end{eqnarray}
With this information, one can derive the key parameters for GW events, including $z$, $d_L$, and $\sigma_{d_L}$. Consequently, we can simulate 1000 GW events expected to be detected by ET during its 10-year observation period \cite{Wang87,Zhang063510,Zhang110431,Zhang068,Jin051,Zhang:2019ple,Li:2019ajo,Zhao:2020ole,Jin:2021pcv,Wu:2022dgy} (see Refs.~\cite{Han:2023exn,Feng:2024lzh} for recent progress).

To estimate the observational constraints on the free parameters of the EMSG models, we employ the MCMC method. The likelihood function for the mock data set of GW standard sirens is formulated as follows
\begin{eqnarray}
\mathcal{L}_{\text{GW}} \propto \text{exp}\left[-\frac{1}{2}\sum_{i=1}^{1000}\left(\frac{d_L^{\text{obs}}(z_i)-d_L^{\text{th}}(z_i)}{\sigma_{d_{L,i}}}\right)\right],
\end{eqnarray}
where the data sets $d_L^{\text{obs}}(z_i)$ denote the luminosity distance of the $1000$ simulated GW events with their associated uncertainties $\sigma_{d_{L,i}}$, while $d_L^{\text{th}}(z_i)$ is the theoretical prediction on each coalescing event. The fiducial values are set to $\Omega_{m}^{\text{fid}}=0.315$ and $H_0^{\text{fid}}=69.8$.

\subsection{Redshift space distortion}

Galaxy surveys are directly sensitive to the combination $f\sigma_8(z)$, which is a nearly model-independent estimator of the observed growth history of the universe. Consequently, many surveys, such as, 2MRS, VVDS, SDSS, 6dFGS, BOSS, Vipers, and so on, offer measurements for this specific estimator. In recent papers \cite{Nesseris023542,Kazantzidis103503}, a compilation of the current measurements for $f\sigma_8(z)$ has been provided, and we utilize these measurements for our analysis.

In order to test the two power-law models described above with the date sets, we use Bayesian inference method. The $\chi^2$ function for the date sets is defined as
\begin{eqnarray}
\chi^2=V^i C_{ij}^{-1}V^j,
\end{eqnarray}
where $V^i=f\sigma_{8,i}-f\sigma_8(z_i;\Theta)$ with $\Theta$ the parameters introduced in the model ($\Theta=H_0,\Omega_{m},\eta_t,\sigma_8$ for $n=1/2$ case and $\Theta=H_0,\Omega_{m},\Omega_Q,\sigma_8$ for $n=1$ case). Here, $f\sigma_8(z_i;\Theta)$ is the theoretical growth function given by the models with a set of parameters $\Theta$ at redshift $z_i$, and $f\sigma_{8,i}$ corresponds to each of the observed data points. Following Ref.~\cite{Kazantzidis103503}, the covariance matrix is constructed as consisting of two parts, i.e., the diagonal and nondiagonal parts. The diagonal part is defined as usual:
\begin{eqnarray}
C_{ii}=\sigma_i^2,
\end{eqnarray}
while the nondiagonal part assesses the potential impact of correlations among different data points. We consider the introduction of randomly selected nondiagonal elements in the covariance matrix while maintaining its symmetry. To be explicit, positive correlations are introduced in 9 randomly selected pairs of data points. The positions of the nondiagonal elements are chosen randomly, and the magnitude of the randomly selected covariance matrix element, denoted as $C_{ij}$, is set according to
\begin{eqnarray}
C_{ij}=0.5\sigma_i\sigma_j,
\end{eqnarray}
where $\sigma_i$ and $\sigma_j$ represent the published 1$\sigma$ errors of the respective data points $i, j$.

The prior is a uniform distribution, where $\Omega_{m} \in [0, 0.5]$, $h \in [0.6, 0.8]$, $\eta_t \in [-1, 1]$, and $\sigma_8 \in [0, 1]$ in the $n=1/2$ case. For the $n=1$ case, the intervals are $\Omega_{m} \in [0, 0.5]$, $h \in [0.6, 0.8]$, $\Omega_Q \in [-1, 1]$, and $\sigma_8 \in [0, 1]$. Besides, the uniform priors are assumed to be the same for the standard sirens data, the RSD data, and the combined data.

\subsection{Analysis}

\begin{figure*}[htb]
\begin{center}
\subfigure[]  {\label{slu}
\includegraphics[width=9cm]{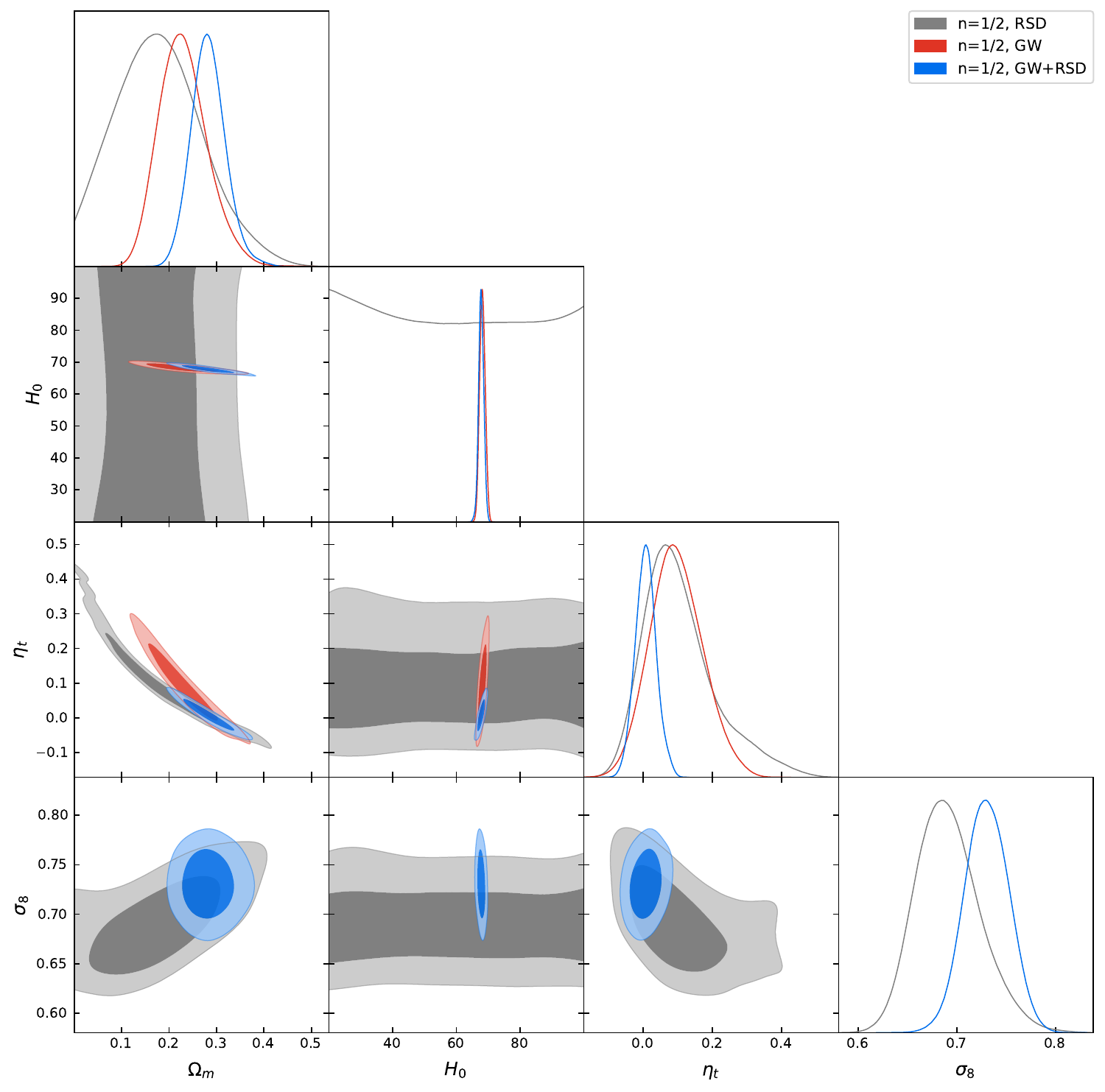}}
\subfigure[]  {\label{slc}
\includegraphics[width=9cm]{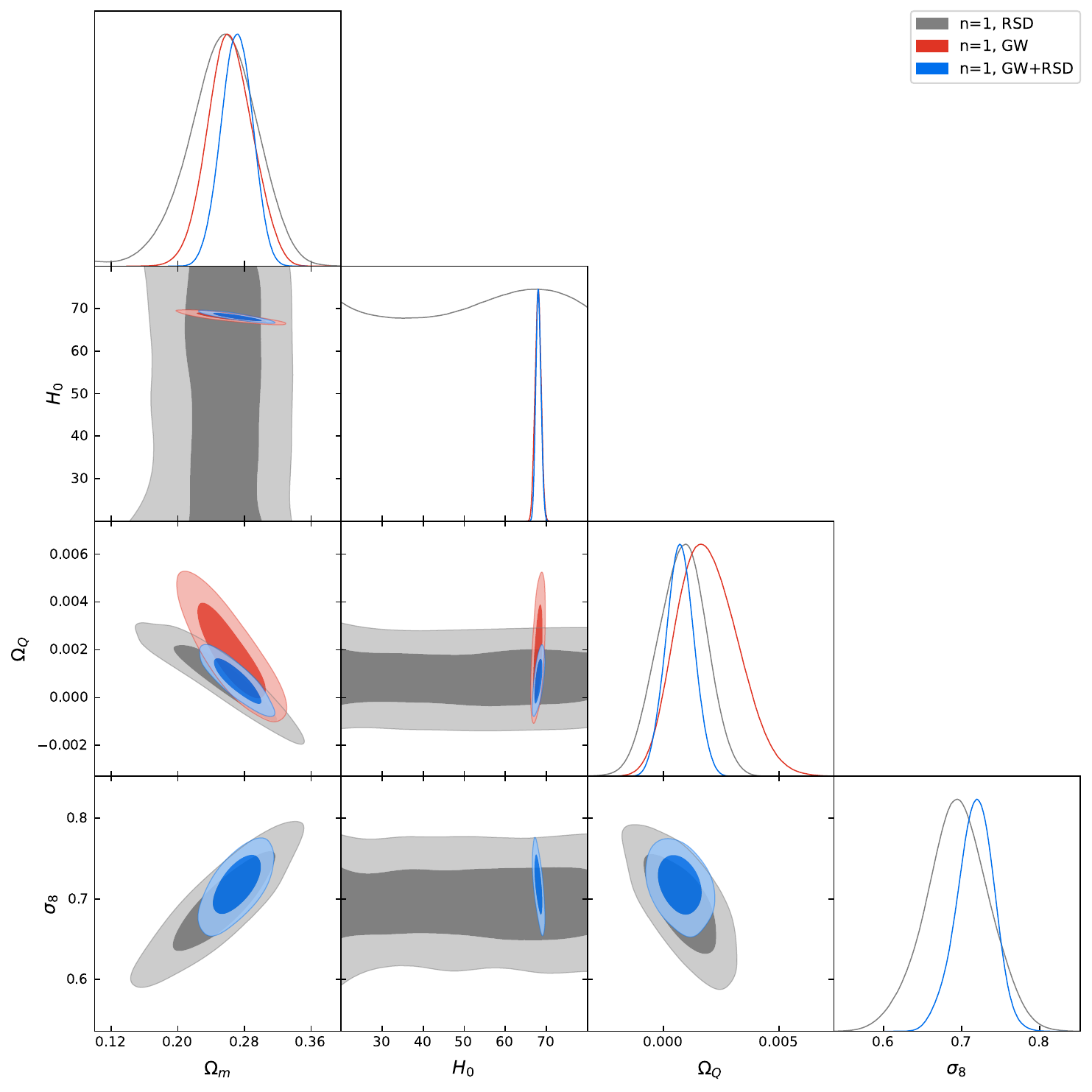}}
\end{center}
\caption{Constraints on the EMSG cosmological model with the RSD and GW data. The upper and lower panel respectively correspond to the posterior distributions of $\Omega_m$, $H_0$, $\eta_t$, and $\sigma_8$ for the case of $n=1/2$ and $\Omega_m$, $H_0$, $\Omega_Q$, and $\sigma_8$ for the case of $n=1$.}
\label{sl}
\end{figure*}

In this subsection, we explore the constraints on the power-law models through a statistical analysis involving three categories of data combinations: GW, RSD, and GW+RSD. We employ the MCMC method for a minimum likelihood $\chi^2$ fit, and the fitting results are summarized in Table \ref{tab1}. The marginalized probability distribution for each parameter ($\Omega_{m}, H_0, \eta_t, \sigma_8$ for the case of $n=1/2$; $\Omega_{m}, H_0, \Omega_Q, \sigma_8$ for the case of $n=1$) along with the marginalized 2D confidence contours are depicted in Figs.~\ref{slu} and \ref{slc}.

Let's begin with the case of $n=1/2$. Figure \ref{sl} presents the fitting results obtained from the RSD data sets, showcasing the fit results: $\Omega_{m}=0.181^{+0.08}_{-0.1}$, $\eta_t=0.106^{+0.064}_{-0.12}$, $\sigma_8=0.691^{+0.027}_{-0.036}$, and no constraint on $H_0$. To facilitate a comprehensive comparison, Fig.~\ref{sl} also includes the best-fit values of the parameters for the simulated GW data are $H_0=68.18\pm 0.86$ km s$^{-1}$ Mpc$^{-1}$, $\Omega_{m}=0.23^{+0.044}_{-0.056}$, and $\eta_t=0.096^{+0.071}_{-0.08}$, {which are nearly determined by the fiducial values as mentioned by many works in the literature \cite{Wang87,Zhang063510,Zhang110431,Zhang068,Jin051}. Even so, the standard siren observation will improve the cosmological parameter estimation since the parameter degeneracy observed in the RSD data differs from that in the standard sirens.} This distinction becomes even more apparent when examining the joint GW+RSD analysis, combining standard probe data sets in the GW and RSD observations. The fit values for the parameters in this combined analysis are $H_0=67.80\pm 0.81$ km s$^{-1}$ Mpc$^{-1}$, $\Omega_{m}=0.283^{+0.033}_{-0.038}$, $\eta_t=0.008\pm 0.03$, and $\sigma_8=0.73\pm 0.022$. Consequently, the shift in the best-fitted parameters, accompanied by a significantly reduced allowed region, demonstrates the potential of leveraging future GW observations to enhance the precision of model parameters in cosmology.

Table \ref{tab1} explicitly presents the best-fit values of the parameters along with their $1\sigma$ uncertainties derived from three distinct data combinations. It's noteworthy that in the realm of modified gravity, the best-fit values for the matter density parameter $\Omega_{m}$ and the Hubble parameter $H_0$ are remarkably consistent with the current estimates within the $\Lambda$CDM cosmology framework, as provided by the Planck data release \cite{AghanimA6}. Furthermore, the deviation parameter $\eta_t$ resulting from the analysis of three data combinations (RSD, GW, and RSD+GW) indicates that the $\Lambda$CDM model ($\eta_t=0$) still falls within a $68.3\%$ confidence level. However, it is worth noting that the best-fit value of $\eta_t$ appears to be slightly larger than $0$, suggesting a potential that the $\Lambda$CDM model might not be the most favored cosmological model according to the current observations.

Turning our attention to the investigation of parameter inference capabilities for the standard siren and RSD in the case of $n=1$, the fit values of the parameters are obtained as {$\Omega_{m}=0.253^{+0.044}_{-0.037}$, $\Omega_Q=0.0008\pm 0.001$, $\sigma_8=0.694\pm 0.041$, and no constraint on $H_0$ for the RSD data set.} The influential role of standard sirens in breaking degeneracies between model parameters is evident, as illustrated by the clearly defined marginalized $1\sigma$ and $2\sigma$ contours for each parameter in Fig.~\ref{slc}. Fitting the $n=1$ model to the combined GW+RSD observations yields parameter values of $H_0=68.06\pm 0.62$ km s$^{-1}$ Mpc$^{-1}$, $\Omega_{m}=0.271\pm 0.019$, $\Omega_Q=0.00072\pm0.0006$, and $\sigma_8=0.717^{+0.025}_{-0.022}$. Table \ref{tab1} summarizes the best-fit values for the three combined datasets. Notably, the deviation from $\Lambda$CDM is also slightly larger than $0$, though still encompassed within a $68.3\%$ confidence level, which also implies that $\Lambda$CDM may not be the most favored cosmological model according to current and future high-redshift GW+RSD observations.

\begin{figure*}[htb]
\begin{center}
\includegraphics[width=8cm]{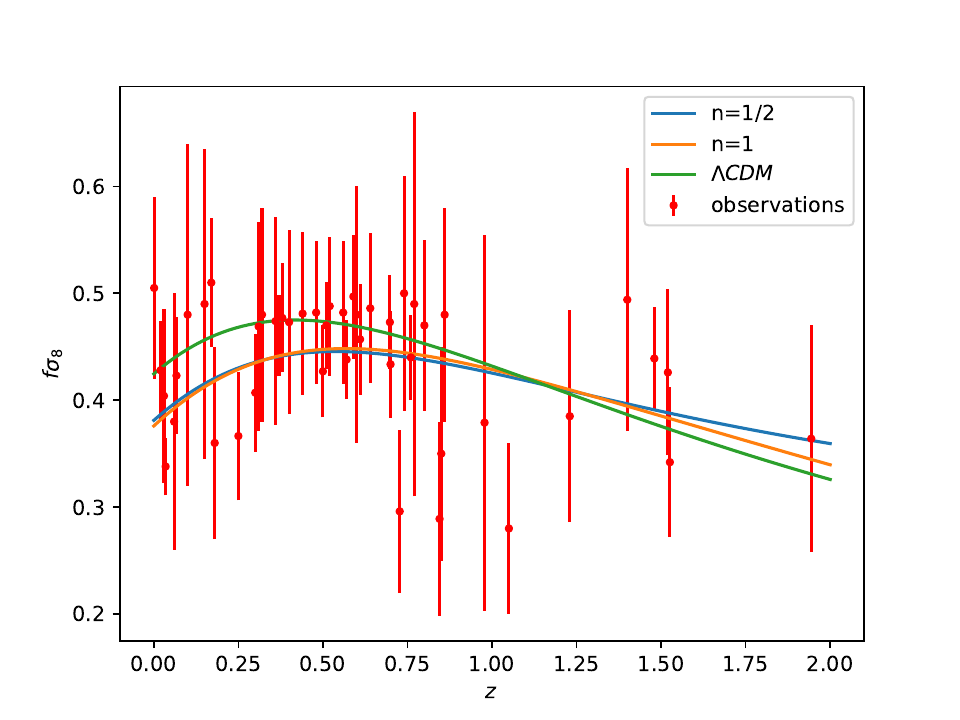}
\end{center}
\caption{The evolution of the $f\sigma_8$ as a function of redshift $z$. Here, the observational data are also shown.}
\label{slp}
\end{figure*}

In Fig.~\ref{slp}, we have illustrated the evolution of $f\sigma_8$ with respect to the redshift. The blue, orange and green lines correspond to the power-law models with $n=1/2$, $n=1$, and $\Lambda$CDM models, respectively. Obviously, the value of $f\sigma_8$ for the cases of $n=1/2$ and $n=1$ is almost the same for lower redshift, both of which are less than that of $\Lambda$CDM model. This indicates that the pow-law models of the EMSG theory predicts the slower growth rate of matter fields. For larger values of the redshift, the evolution of the $f\sigma_8$ for the case of $n=1/2$ decreases slower than the case of $n=1$ and $\Lambda$CDM models. Thus, more observations would be needed to favor one of them.

Finally, we need to make some critical clarifications. In this paper, the RSD data we used are real observational data, but the GW data are simulated. The RSD data are entirely incapable of constraining the Hubble constant, resulting in a clear parameter degeneracy within the parameter space. The GW data can measure the Hubble constant very well, so when the RSD data are combined with the GW data, the original degeneracy is broken, and the improvement in parameter constraints is very significant. Since the GW data are simulated data, the central values of the cosmological parameters are fixed in the simulation; thus, in our final joint fit results, the significance of the central values is not particularly great, and the most meaningful result is the range of errors. What our work aims to demonstrate is precisely this point: future GW observations, if combined with RSD observations to constrain EMSG theory, can greatly enhance the precision of cosmological parameter constraints.

\begin{table*}
\begin{center}{\scriptsize
\caption{The marginalized $1\sigma$ uncertainties of the parameters
for the two viable models in EMSG theory. Note that here $H_0$ is in units of km s$^{-1}$ Mpc$^{-1}$.}~\label{tab1}
\begin{tabular}{|c|c|c|c|c|} \hline\hline
\cline{1-5}
Data  \ \ & \ \ $\Omega_{m}$   &$H_0$   &$\eta_t$   &$\sigma_8$ \\ \hline
GW+RSD\ \ & \ \ $0.283^{+0.033}_{-0.038}$ \ \ & \ \ $67.80\pm 0.81$\ \ & \ \ $0.008\pm 0.03$\ \ & \ \ $0.73\pm 0.022$\ \\ \hline
GW\ \  & \ \ $0.23^{+0.044}_{-0.056}$ \ \ & \ \ $68.18\pm 0.86$ \ \ & \ \ $0.096^{+0.071}_{-0.08}$\ \ & \ \ $-$\ \\ \hline
RSD\ \ & \ \ $0.181^{+0.08}_{-0.1}$ \ \ & \ \ $-$ \ \   & \ \ $0.106^{+0.064}_{-0.12}$\ \ & \ \ $0.691^{+0.027}_{-0.036}$\ \\
\hline \hline
Data  \ \ & \ \ $\Omega_{m}$   &$H_0$   &$\Omega_Q$   &$\sigma_8$ \\ \hline
GW+RSD\ \ & \ \ $0.271\pm 0.019$ \ \ & \ \ $68.06\pm 0.62$\ \ & \ \ $0.00072\pm 0.0006$\ \ & \ \ $0.717^{+0.025}_{-0.022}$\ \\ \hline
GW\ \  & \ \ $0.263\pm 0.027$ \ \ & \ \ $67.98\pm 0.72$ \ \ & \ \ $0.0019^{+0.0012}_{-0.0014}$\ \ & \ \ $-$\ \\ \hline
RSD\ \ & \ \ $0.253^{+0.044}_{-0.037}$ \ \ & \ \ $-$ \ \   & \ \ $0.0008\pm 0.001$\ \ & \ \ $0.694\pm 0.041$\ \\
\hline\hline
\end{tabular}}
\end{center}
\end{table*}

\section{Conclusion}~\label{con}

In this paper, we developed the theory of the linear cosmological perturbations for the energy momentum squared tensor gravity theory. We mainly focused on the scalar and tensor modes since the vector mode decays rapidly with the cosmological expansion. For the tensor perturbation, we showed that the propagation speed of gravitational wave is the same as the light which is in agreement with the recent multimessenger measurements. Regarding scalar perturbation, our focus has been exclusively on the sub-horizon limit. This approach aims to elucidate the equations governing the linear formation of cosmological structures, essential for assessing these theories against existing cosmological precision data. Notably, these equations have been absent in the existing literature, and our investigation seeks to fill this gap.

Then, by utilizing the simulated data generated by the third-generation GW detector, i.e., the Einstein Telescope, we explored the constraining capabilities of GW events on the EMSG theory. We mainly focused on two specific pow-law models, serving as significant viable models for cosmology. Our findings demonstrate that the sensitivity achievable by the ET detector or a comparable third-generation interferometer is sufficient to enhance the current parameter estimates within the EMSG theory. Besides, we also explored the constraining capabilities of the growth of matter density perturbation with the RSD data sets. Owing to the non-conservative nature of the theory, the evolution equation for the matter density perturbation undergoes modification within these models. We found that the RSD data sets alone may not offer the constraints comparable to the GW observations. However, when combined with the standard sirens, the precision of the constraint is largely imporved since the degeneracies among parameters are mitigated. Although the $\Lambda$CDM model (corresponding to $\eta_t=0$ or $\Omega_Q=0$) still falls within a $68.3\%$ confidence level, the nonvanishing of these two modified gravitational parameters indicates that the $\Lambda$CDM model might not be the most favored cosmological model according to the current observations. The plot of $f\sigma_8$ demonstrates that, for lower redshifts where the data availability is higher, the pow-law models of the EMSG theory predicts a slower growth rate of matter fields. While, for larger redshifts, the evolution of $f\sigma_8$ for the pow-law models decreases slower than the prediction of the $\Lambda$CDM model. Thus, more observations would be needed to favor one of them.

\begin{widetext}

\appendix
\section~\label{A}

\begin{eqnarray}
A(t)&=&\frac{2}{\left(1+2 \rho  f_{\mathbb{T}}\right) \left(1+2 \rho  f_{\mathbb{T}}+4 \rho ^3 f_{\mathbb{T}\mathbb{T}}\right)}\bigg(H \left(1+\rho  \left(f_{\mathbb{T}}-2 \rho  f_{\mathbb{T}}^2-8 \rho ^2 f_{\mathbb{T}\mathbb{T}}-28 \rho ^3 f_{\mathbb{T}}f_{\mathbb{T}\mathbb{T}}\right)\right) \nonumber\\
&+&\rho\bigg(2 \rho  \left(2 \rho  \left(1+2 \rho  f_{\mathbb{T}}\right) \dot{f}_{\mathbb{T}\mathbb{T}}+\left(3+4 \rho  f_{\mathbb{T}}\right) f_{\mathbb{T}\mathbb{T}} \dot{\rho}\right)-\frac{\dot{f}_{\mathbb{T}} \left(4 \rho ^2 \left(f_{\mathbb{T}} \left(f_{\mathbb{T}}+10 \rho ^2 f_{\mathbb{T}\mathbb{T}}\right)+4 \rho  f_{\mathbb{T}\mathbb{T}}\right)-1\right)}{1+2 \rho  f_{\mathbb{T}}}\bigg)\bigg),
\end{eqnarray}
\end{widetext}
\begin{widetext}
\begin{eqnarray}
B(t)&=&\frac{1}{2 \left(1+2 \rho  f_{\mathbb{T}}\right) \left(1+2 \rho  f_{\mathbb{T}}+4 \rho ^3 f_{\mathbb{T}\mathbb{T}}\right)}\bigg(-\frac{\rho  \left(1+2 \rho  f_{\mathbb{T}}\right)^2 \left(1+2 \rho  f_{\mathbb{T}}+4 \rho ^3 f_{\mathbb{T}\mathbb{T}}\right) \left(a^2 f_R+4 k^2 f_{RR}\right)}{2 f_R \left(a^2 f_R+3 k^2 f_{RR}\right)} \nonumber\\
&+&4\bigg(\dot{\rho} \dot{f}_{\mathbb{T}}-24 H \rho ^3 f_{\mathbb{T}\mathbb{T}} \left(H+2 H \rho  f_{\mathbb{T}}+\rho  \dot{f}_{\mathbb{T}}\right)-12 H \rho ^3 \left(1+2 \rho  f_{\mathbb{T}}\right) \dot{f}_{\mathbb{T}\mathbb{T}}+6 \rho ^2 \dot{\rho} \dot{f}_{\mathbb{T}\mathbb{T}}+16 \rho ^3 f_{\mathbb{T}} \dot{\rho} \dot{f}_{\mathbb{T}\mathbb{T}} \nonumber\\
&-&8 \rho ^4 \dot{f}_{\mathbb{T}} \dot{f}_{\mathbb{T}\mathbb{T}}+2 H \rho  \left(\dot{f}_{\mathbb{T}}+2 \rho ^2 \left(1+2 \rho  f_{\mathbb{T}}\right) \dot{f}_{\mathbb{T}\mathbb{T}}\right)+\rho  \ddot{f}_{\mathbb{T}}-12 \rho ^2 f_{\mathbb{T}\mathbb{T}} \Big(H \left(\dot{\rho} \left(3+8 \rho  f_{\mathbb{T}}\right)+2 \rho ^2 \dot{f}_{\mathbb{T}}\right) \nonumber\\
&+&\rho  \left(\left(1+2 \rho  f_{\mathbb{T}}\right)\dot{H} +4 \dot{\rho} \dot{f}_{\mathbb{T}}+\rho  \ddot{f}_{\mathbb{T}}\right)\Big)+2 \rho ^3 \left(1+2 \rho  f_{\mathbb{T}}\right) \ddot{f}_{\mathbb{T}\mathbb{T}}+\frac{k^2 \rho  f_{\mathbb{T}} \left(1+2 \rho  f_{\mathbb{T}}\right)}{a^2} \nonumber\\
&+&\frac{48 \rho ^3 f_{\mathbb{T}\mathbb{T}} }{1+2 \rho  f_{\mathbb{T}}}\left(f_{\mathbb{T}} \dot{\rho}+\rho  \dot{f}_{\mathbb{T}}\right) \left(H+2 H \rho  f_{\mathbb{T}}+\rho  \dot{f}_{\mathbb{T}}\right)+\frac{24 \rho ^4 f_{\mathbb{T}} f_{\mathbb{T}\mathbb{T}} }{\left(1+2 \rho  f_{\mathbb{T}}\right)^2}\left(H+2 H \rho  f_{\mathbb{T}}+\rho  \dot{f}_{\mathbb{T}}\right) \left(H \left(3+6 \rho  f_{\mathbb{T}}\right)+2 \rho  \dot{f}_{\mathbb{T}}\right) \nonumber\\
&-&\frac{4 \rho  }{1+2 \rho  f_{\mathbb{T}}}\left(f_{\mathbb{T}} \dot{\rho}+\rho  \dot{f}_{\mathbb{T}}\right) \left(\dot{f}_{\mathbb{T}}+2 \rho ^2 \left(1+2 \rho  f_{\mathbb{T}}\right) \dot{f}_{\mathbb{T}\mathbb{T}}\right) \nonumber\\
&-&\frac{2 \rho ^2 f_{\mathbb{T}} }{\left(1+2 \rho  f_{\mathbb{T}}\right)^2}\left(H \left(3+6 \rho  f_{\mathbb{T}}\right)+2 \rho  \dot{f}_{\mathbb{T}}\right) \left(\dot{f}_{\mathbb{T}}+2 \rho ^2 \left(1+2 \rho  f_{\mathbb{T}}\right) \dot{f}_{\mathbb{T}\mathbb{T}}\right)\bigg)\bigg).
\end{eqnarray}
\end{widetext}

\acknowledgments{
This work was supported by the National SKA Program of China (Grants Nos. 2022SKA0110200 and 2022SKA0110203), the National Natural Science Foundation of China (Grants Nos. 11975072, 11875102, 11835009 and 12247101), and the National 111 Project (Grant No. B16009).
}

\end{document}